# ACCOMPLISHMENTS OF THE HEAVY ELECTRON PARTICLE ACCELERATOR PROGRAM*

D. Neuffer†, D. Stratakis, Fermilab, Batavia IL 60510, USA, M. Palmer, BNL, Upton, NY 11973, J-P Delahaye, SLAC, Menlo Park, CA 60439, USA, D. Summers, U. Miss., Oxford MS 38655, USA, R. Ryne, LBNL, Berkeley CA 94720, USA, M. A. Cummings, Muons, Inc.

*Abstract*

The Muon Accelerator Program (MAP) has completed a four-year study on the feasibility of muon colliders and on using stored muon beams for neutrinos. That study was broadly successful in its goals, establishing the feasibility of heavy lepton colliders (HLCs) from the 125 GeV Higgs Factory to more than 10 TeV, as well as exploring using a μ storage ring (MSR) for neutrinos, and establishing that MSRs could provide factory-level intensities of $\nu_e$ ($\bar{\nu}_e$) and $\bar{\nu}_\mu$ ($\nu_\mu$) beams. The key components of the collider and neutrino factory systems were identified. Feasible designs and detailed simulations of all of these components have been obtained, including some initial hardware component tests, setting the stage for future implementation where resources are available and the precise physics goals become apparent.

## INTRODUCTION

Initial concepts for muon colliders and muon storage rings were proposed in ~1980[1-4], and research toward these concepts intensified in the 1990's in the search for feasible high-energy accelerator projects. In 2011, muon accelerator R&D in the United States was consolidated into a single entity, the Muon Accelerator Program (MAP) [5]. The purpose of MAP was to perform R&D in heavy electron (μ) accelerator technologies and to perform design & simulation to demonstrate the *feasibility* of concepts for neutrino factories and muon colliders [6,7,8]. MAP established that general feasibility, and awaits the development of physics motivations before proceeding to full implementation. The design studies have been accompanied by technology R&D, establishing the feasibility of key scenario components. Though MAP did not include detailed engineering studies, the design studies were performed with an awareness of gradient and field limits, and space requirements for hardware, etc. The following highlights some key accomplishments under MAP in design concepts for muon-based accelerators for neutrino factories and muon colliders.

## DESIGN OVERVIEW

The key components of collider and neutrino factory systems were identified and are displayed in block diagram form in Figure 1. These are a high-intensity proton source, a multi-MW target and transport system for π capture, a front end system for bunching, energy compression and initial cooling of μ's, muon cooling systems to obtain intense μ⁺ and μ⁻ bunches, acceleration up to multiTeV energies, and a collider ring with detectors for high luminosity collisions. For a neutrino factory a similar system could be used but with a racetrack storage ring for ν production and without the cooling needed for high luminosity collisions. The Proton Driver, Target, Front End, and Acceleration linac, are common to both facilities.

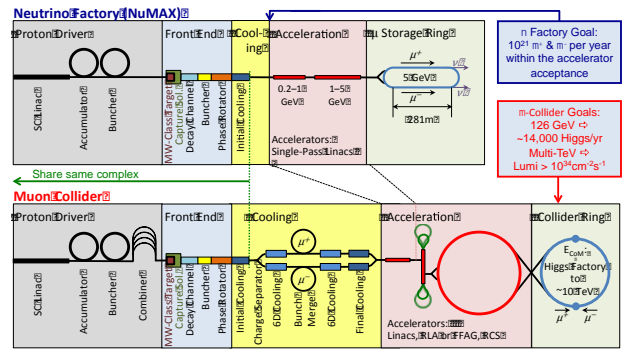

Figure 1: Block diagram of neutrino factory and muon collider facilities, as studied under MAP.

## PROGRESS IN MUON ACCELERATOR DESIGN UNDER MAP

Though MAP existed for only 4 years, there has been tremendous progress in the design concepts. Some highlights include:

*Proton Driver:* Under MAP, designs were developed for the accumulator and compressor rings of the Proton Driver, based on the expected parameters of the Project-X linac [9]. Potential instabilities were analyzed and initial studies were performed of injection stripping and of the beam to target delivery system for the HLC design. Meanwhile, JPARC has directly demonstrated that a proton source can operate at MAP-required parameters. A proton driver based on a JPARC-style linac + rapid-cycling synchrotron is an attractive possibility [10].

*Target & Front End:* MAP has explored several target designs, including a design based on a solid carbon target and on a liquid Mercury target [11, 12]. The target parameters have been optimized [13]. The Front End designs use a novel rf buncher and phase-energy rotator to form the beam into a train of μ⁺ and μ⁻ bunches that can be cooled, and accelerated by downstream systems [13, 14]. An energy deposition control system using a chicane and downstream absorber was also invented [15, 16].

*Cooling:* Muon cooling designs matured greatly under MAP. Figure 2 shows how the horizontal and vertical emittances evolve as the muons travel through the cooling subsystems. When MAP began there was not an accepted approach to how the various subsystems should be organized. Under MAP, start-to-end simulations have now been performed of vacuum and gas-filled cooling systems to reach the minmal emittance (see Fig. 3) [17-19]. These start with a "FOFO snake" cooling section, which can cool both $\mu^+$ and $\mu^-$ simultaneously [20, 21]. This is followed by a 6D cooling system, a bunch merge [22, 23], and a post-merge 6D cooling system. An important development under MAP, discovered by Balbekov, is that 6-D cooling can, be achieved using a rectilinear channel with slightly tilted solenoids and does not require large-aperture bending magnets[24, 25]. Under MAP there have been major advances in the design & simulation of a gas-filled Helical Cooling Channel (HCC) [26, 27]. The HCC is compact and can tolerate high gradient RF in magnetic fields by the use of gas-filled cavities. The rectilinear channel can also use gas-filled rf [28, 29]. A final emittance exchange to minimal transverse emittance is needed for a muon collider [30,31].

---

\* supported by Fermi Research Alliance, LLC under Contract No. De-AC02-07CH11359 with the U. S. Department of Energy.
† neuffer@fnal.gov

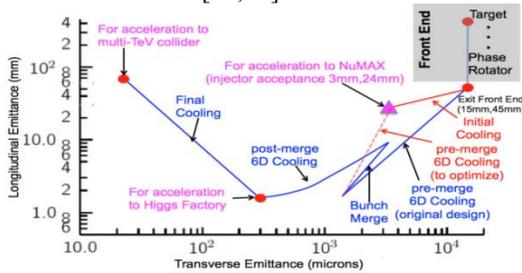

Figure 2: Transverse and longitudinal emittance evolution in a muon cooling system.

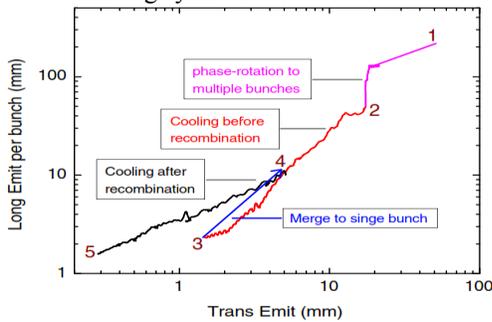

Figure 3: In a key accomplishment of the MAP program, cooling systems were designed and simulated that can provide all of the cooling needed for a collider, using feasible magnet and rf designs.

*Acceleration:* Under MAP, it was shown that, for low energies (up to ~ 5 GeV), a dual-use linac accelerating both proton and muon beams is a viable option [32]. Multi-pass recirculating linear accelerators (RLAs) are an efficient means of acceleration up to a few 10's of GeV, as needed for a Higgs Factory[33], and could also be used for higher energies . Hybrid rapid-cycling synchrotrons, containing ramped normal conducting magnets and fixed-field SC magnets, were designed and could be more economical for acceleration from ~100 GeV to the multi-TeV range [34]. Fast-ramping magnets suitable for the RCS were designed and tested by Piekarz et al.[35]

*Collider Rings:* Under MAP, collider ring designs were developed for a Higgs Factory, and for 1.5 TeV, 3 TeV, and 6 TeV colliders [36-37]. These took into account many factors including the design of magnets able to survive in the environment of a stored muon beam, the design of final focus systems, halo extraction, longitudinal dynamics including wakefield effects and chromaticity correction, and beam-beam effects.

A critical feature of the rings is that the muon beam energy can be measured very accurately by tracking the stored beam spin precession (as is done for the g-2 experiment)[38].

*Machine-Detector Interface (MDI):* Many improvements were made to MARS15, the leading particle interaction program, and applied to MAP. MARS was used for many purposes across the full range of MAP designs, including target studies, component and detector shielding studies, and background simulations for detector studies for colliders [39,40].

*Muon Decay Rings:* Under MAP, designs were developed for a short-baseline neutrino facility (nuSTORM) and a long-baseline neutrino Factory (NuMAX) [41-43, 6]. The nuSTORM design used MAP concepts to develop a modest μ storage ring that could test for sterile ν's, measure ν cross sections and provide low-E μ beams for cooling and other experiments. The NuMAX design would extend the DUNE experiment with a high-intensity ν-factory for complete ν-oscillation measurements.

*High-End Computing:* Prior to MAP most simulations were performed with serial codes. Particle simulations typically used at most 100,000 particles, often less, and in some cases required many hours to run. The main codes used for design & simulation at MAP were G4Beamline, ICOOL, and MARS. Under MAP, ICOOL and G4Beamline were parallelized. All three codes were installed at the NERSC supercomputer. Also, the SPACE code was developed to simulate the interaction of intense beams with plasmas in HPRF cavities [44]. Parallel scans with capabilities for parallel design optimization were developed, including a Genetic Algorithm for magnetic horn optimization for NuSTORM.[45]

*Low-energy Muon Applications:* Prior to MAP, the neutrino factory and muon collider collaboration made critical contributions initiating the mu2e and g-2 experiments at Fermilab. These contributions have continued as these projects have initiated construction. Further R&D based on MAP can provide the basis for higher-intensity upgrades of these experiments, or other experiments exploring lepton parameters.

*High-field Magnet Development:* HLC performance depends directly on magnetic field. The MAP program included designs and initial tests on high field magnets, with $Nb_3Sn$ and HTS conductors, as well as NbTi designs. [46]

*Rf Development:* At the time MAP was initiated there was significant concern that RF cavities could not operate at high gradients with the focusing magnetic fields. Under MAP these phenomena has been understood and several solutions demonstrated. Careful cavity design enables higher gradients with increasing magnetic field. Beryllium has been shown to have almost no damage due to breakdown compared with copper. Experiments at the Fermilab MuCool Test Area (MTA) have demonstrated that using cavities filled with high-pressure gas can prevent this breakdown; and this is a viable technology for muon cooling [47, 48].

*International Muon Ionization Cooling Experiment (MICE).* MICE, based at RAL (UK), is an international experiment to test ionization cooling and MAP is a major contributor [49]. MICE has developed and demonstrated the capability for precision measurements of μ beam before and after a cooling segment [50]. It has or will test key components of a cooling system, including $H_2$ and LiH absorbers, magnets, rf, and emittance exchange.

## CONCLUSION

The design & simulation work and technology R&D done under MAP made significant advances in demonstrating the feasibility of muon accelerators. Under MAP, key technological obstacles have been overcome (e.g., high gradient RF in magnetic fields, and development of 6-D cooling scenarios). MAP designers demonstrated via simulation the performance of realistic system designs for a neutrino factory and nearly all sub-systems required for a muon collider.

An important prerequisite for a High Energy Heavy Lepton Collider (HLC) is a multi-MW-scale proton source, as could be developed at JPARC or ESS; however, the US HEP program does not have one. Since feasibility has been established by MAP and detailed implementation cannot begin until a proton source is established, it could be expedient to focus accelerator resources on initiating the proton source and defer an ambitious collider program.

Within the limited US high-energy physics budget and project constraints, the largest initiative that the 2014 HEPAP panel could envision for the next decade is a deep underground neutrino experiment. Initiation of a high intensity proton source was included in that program. MAP research efforts were curtailed, having successfully completed the feasibility assessment.

Critical research important for a future collider is nonetheless continuing, outside the MAP framework. The 2014 HEPAP panel supported high-field magnet development, which is critical for future HLCs, since beam production, beam cooling, acceleration and collider performance directly depend on the magnetic field strength. Optimization of technology for secondary particle production is a HEPAP priority, as is also rf gradient increases. The g-2 and μ2e experiments at Fermilab will provide important experience in using and optimizing μ beams, including precision spin precession measurements.

While this supporting technology R&D is helpful, some dedicated research on HEPA will be needed to maintain its availability for future accelerators. This research should be internationally based, since any future HEP facility will require international support and the US HEP program may not have the resources for a next generation facility. This places increased importance on international collaboration, such as the UK-based MICE effort, which is the only remaining funded activity.

This research should be enlightened by the changing landscape in HEP. At present, ν experiments are focused on using π-decay $ν_μ$-beams to measure the parameters of the 3-ν standard model, with the next experiments to determine the mass hierarchy and to determine CP violation at the ~5σ level, if it be near maximal. If the goal after that is greater accuracy, MAP has established that a μ-accelerator based ν-beam could do this. If the ν physics is more complex, with more ν's or unexpected interactions, then it is probable that ν-beams from μ acceleration and storage will be needed. Since the facility needed for further exploration after 2030 may differ substantially from the present concepts, a renewed design and optimization effort is essential for a healthy HEP program.

A muon accelerator facility also holds significant promise for precision capabilities spanning the Intensity and Energy Frontiers, including precision symmetry experiments (following μ2e, …) as well as the HE frontier.

LHC with its extensions to higher luminosity and maximal energy is the current HEP discovery machine. So far, its novel discoveries are limited to the Higgs at 125 GeV and the absence of new HE particles beyond that. A primary purpose of a lepton collider is detailed exploration of established or expected resonance states (J/ψ, Υ, $Z_0$, …); identification of any at higher energy by LHC or theoretical physics would motivate the construction of a HLC.

If more precise measurements of the Higgs properties are needed, in particular measurements of its mass, width, and its coupling to second generation leptons, then a 125 GeV $μ^+$-$μ^-$ collider would provide the highest precision. Since μ beam energies can be measured by spin precession (frequency), rather than by calorimetry or bending radius, they can be measured much more accurately. Masses and widths of the nearby $Z_0$ and tt* resonances could also be measured, completing a precision scan of the standard model at highest possible accuracy [51].

The absence of new HE particles may indicate the need for a higher energy machine. A ~10TeV HLC could have the discovery reach of a 100+ TeV pp collider, and could be considered if the cost and scale of a hadron collider becomes unacceptable.